# A Planar Tracking Game with Sensing Delays and its MATLAB Implementation

Nan Yu[1], Chifu Yang[2], Miao Li[3]

[1]University College Dublin, Dublin, D04 V1W8, Ireland
[2]Harbin Institute of Technology, Harbin, 150006, China
[3]Nagoya University, Nagoya, 465-8601, Japan

**Abstract:** This paper proposes a new perspective on the conventional planar target tracking problem. One evader and one pursuer are considered in the dynamics. In the planar tracking, pursuer has the ability to measure the position and the velocity information of the evader but with sensing delays. The modeling and the controller design of the system are presented with details. Then, a computer game is developed and implemented using MATLAB/Simulink, which constitutes the main contribution of the paper.

*Keywords*: Tracking game, Delay, MATLAB, Simulink

## 1. Introduction

The target tracking problem has various real-world applications, such as surveillance, robotic navigation, robotic manipulator operation, and torpedo guidance [1-6]. The goal of trajectory tracking control of robots is to control the position and orientation of manipulators, which is indispensable for industrial applications [7-11]. Sometimes the control goal is a fixed point and other times the control goal is a continuous time-variant track. In this paper, we consider a classic pursuit-evasion situation, also known as target tracking problem, with a novel perspective. On a 2D plane, a pursuer tries to capture an evader using the real-time feedback control law. Along this feedback loop, signals are processed and transmitted, yielding time delays in processing and transport. Delays are ubiquitously observed in real-world systems due to the unavoidable time required to gather sensing data needed for decision-making, to generate control decisions, and to execute these decisions [12-15]. Here we only consider the sensing delay which is of paramount significance compared to other delays. Due to the delayed sensory perception on the pursuer side, the capture may or may not be guaranteed. In fact, these sensing delays are crucial factors that may deteriorate or even destabilize the performance of the control systems [16-19].

The trajectory of the evader is taken as totally arbitrary and can be freely controlled by a human operator. This trajectory is unknown to the pursuer. We consider two time delays in the dynamics, one in the perception of the position information of the evader and the other in the sensing of the velocity information of the evader. With these, the overall dynamics could be modeled as a linear time invariant multiple time delay systems (LTI-MTDS). The trajectory tracking problem with time delayed feedback has received a great deal of attention in the control systems area. A recent paper [2] uses a Lyapunov-based adaptive control design to achieve the global stability of a general class of plants, under time-delayed state feedback. Reference [4] deals with a similar problem, using feed-forward adaptive control to compensate against the delay. The Lyapunov-Krasovskii functional method is deployed to assure stability. A globally stable adaptive controller with a Smith-predictor- like structure for SISO input-delayed plants was analyzed in [7].

Many control treatments encountered in literature resolve this problem by stabilizing the delayed dynamics using forecasting and state estimation methods and deploying the estimated states in the control law instead of their time-delayed forms. Similar treatments have been proven successful in many applications to the cost of numerical overhead and the drawbacks of uncertainty carried along with the adaptive procedures [1, 3, 20-29].

## 2. Modeling and Control

Both the pursuer and the evader are considered as point masses moving on a 2-dimensional plane. The position vectors of the pursuer and the evader are denoted by $r_p = [x_p(t) \ y_p(t)]^T$ and $r_e = [x_e(t) \ y_e(t)]^T$, respectively. The pursuer is the only controlled agent, and its dynamics is governed by

$$m\ddot{r}_p(t) = -c\dot{r}_p(t) + u_p(t) \tag{1}$$

where $c$ is the viscous damping coefficient between the pursuer and the operating platform, and $u_p(t)$ is the vector control input. Considering the state vector $z_p = [x_p \ \dot{x}_p \ y_p \ \dot{y}_p]^T$, this dynamics can be represented in the state space form as

$$\dot{z}_p(t) = A_p z_p(t) + Bu(t) \tag{2}$$

$$\mathbf{A}_p = \begin{bmatrix} 0 & 1/m & 0 & 0 \\ 0 & -c/m & 0 & 0 \\ 0 & 0 & 0 & 1/m \\ 0 & 0 & 0 & -c/m \end{bmatrix}, \quad \mathbf{B} = \begin{bmatrix} 0 & 0 \\ 1/m & 0 \\ 0 & 0 \\ 0 & 1/m \end{bmatrix}$$

The evader state, represented by $z_e = [x_e \ \dot{x}_e \ y_e \ \dot{y}_e]^T$, follows a trajectory which is unknown a priori and is determined by the human operator. In fact, in the computer game to be developed using MATLAB in the next section, this trajectory is dictated by the human operator moving a mouse cursor. The evader follows a filtered version of the trajectory of a mouse pointer in order to limit its speed. The filter is designed to guarantee no overshoot and a reasonable settling time.

We define the error vector of the pursuer as:

$$e(t) = z_p(t) - z_e(t) \tag{3}$$

and the resulting error dynamics can be written as

$$\dot{e}(t) = A_p e(t) + Bu(t) \tag{4}$$

For the control input in (4), we use the full state feedback, $u = -ke(t)$, where the gain matrix $k \in \Re^{2 \times 4}$ is designed as an optimal LQR controller for the non-delayed system.

The synthesized control logic is then applied to the dynamics of the pursuer, leading to:

$$\dot{z}_p(t) = A_p z_p(t) - Bke(t) \tag{5}$$

Now, we propose a new perspective to the problem at hand. Assume that due to cognitive delays on the pursuer side, the pursuer is unable to measure the current location and velocity of the evader. Denote that the delay in the position measurement is $\tau_1$ seconds and another delay of $\tau_2$ seconds in the processing of velocity information. Considering the effects of the time delay, the new dynamics becomes:

$$\dot{z}_p(t) = A_p z_p(t) + B_1 e(t - \tau_1) + B_2 e(t - \tau_2) \tag{6}$$

where the matrices affecting the delay terms are:

$$\mathbf{B}_1 = \begin{bmatrix} 0 & 0 & 0 & 0 \\ -k_{11} & 0 & 0 & 0 \\ 0 & 0 & 0 & 0 \\ 0 & 0 & -k_{23} & 0 \end{bmatrix}, \quad \mathbf{B}_2 = \begin{bmatrix} 0 & 0 & 0 & 0 \\ 0 & -k_{12} & 0 & 0 \\ 0 & 0 & 0 & 0 \\ 0 & 0 & 0 & -k_{24} \end{bmatrix} \tag{7}$$

In fact, the stability of the pursuer dynamics is defined by the homogeneous solution of (6). The characteristic equation of this system is given by:

$$\det(s\mathbf{I} - \mathbf{A}_p - \mathbf{B}_1 e^{-\tau_1 s} - \mathbf{B}_2 e^{-\tau_2 s}) = 0 \tag{8}$$

To study the stability of (8) for all possible delay combinations is not a trivial task. The difficulty is created by the infinite dimensionality which is introduced by the transcendental terms in (8). Many theories on time-delay systems could be utilized to assess the overall stability of (8) [30-39]. The focus of this paper is on the implementation of the target tracking game using MATLAB/Simulink, which is elaborated on in the next section.

## 3. MATLAB/Simulink Implementation

The interactive computer game is developed and implemented in MATLAB/Simulink. Figure 1 shows the main diagram with detailed illustration for each component.

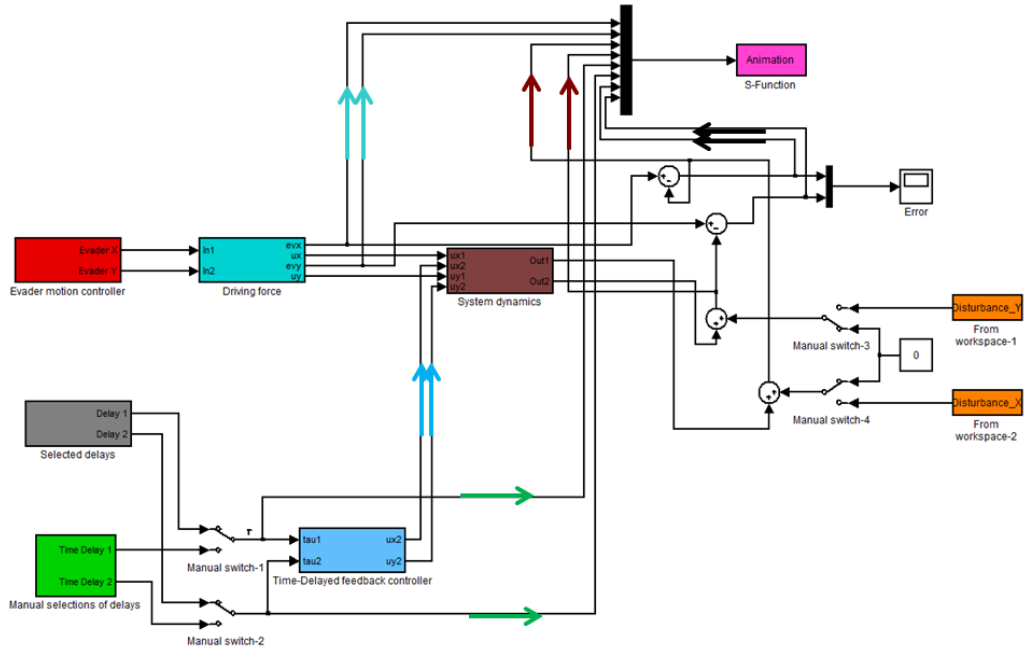

**Figure 1.** Complete control chart--red: evader motion controller, cyan: driving force, brown: system dynamics, grey: selected delays, green: manual selections of time delays, blue: time-delayed feedback controller, orange: disturbances in X and Y direction, pink: S-Function (animation)

The evader motion controller is the mechanism which drives the evader to the cursor (driven by the player via a mouse input). It is composed of a mouse reader and two decoupled 1-D motion controllers as in Figure 2.

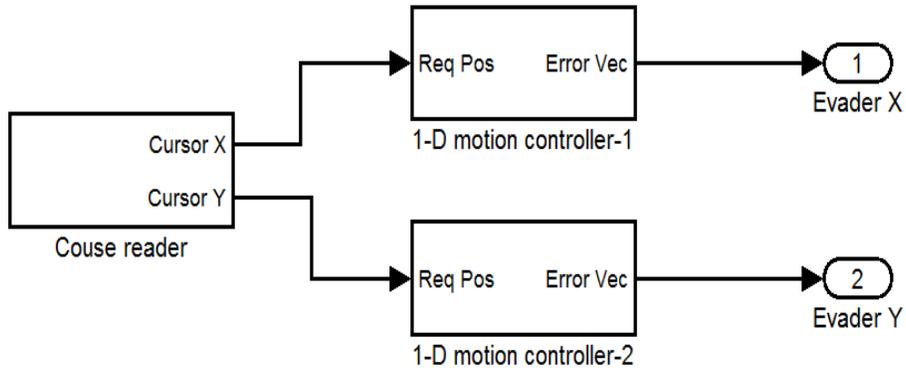

**Figure 2.** Evader Motion Controller

The cursor reader gets the position coordinates of the cursor. The 1-D motion controller given in the Figure 3 drives the evader to track the cursor. It's a unity feedback control with a PI controller. The characteristic roots of the transfer function between the position of the evader and that of the cursor are real.

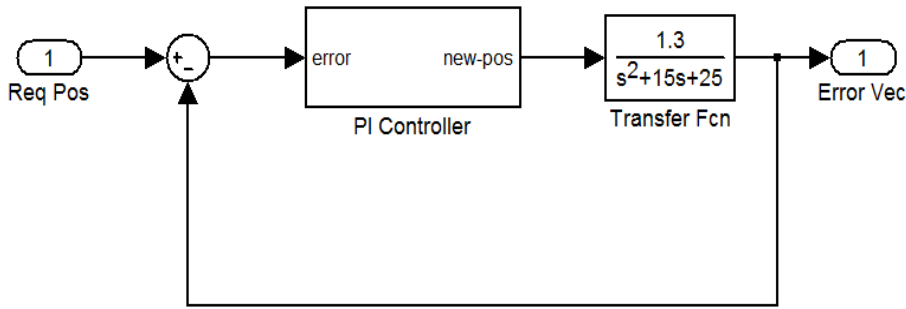

**Figure 3.** 1-D motion controller for the evader (Required Position comes from Cursor X/Cursor Y)

Driving force block (Figure 4) simply creates the position control command for the evader in X and Y direction.

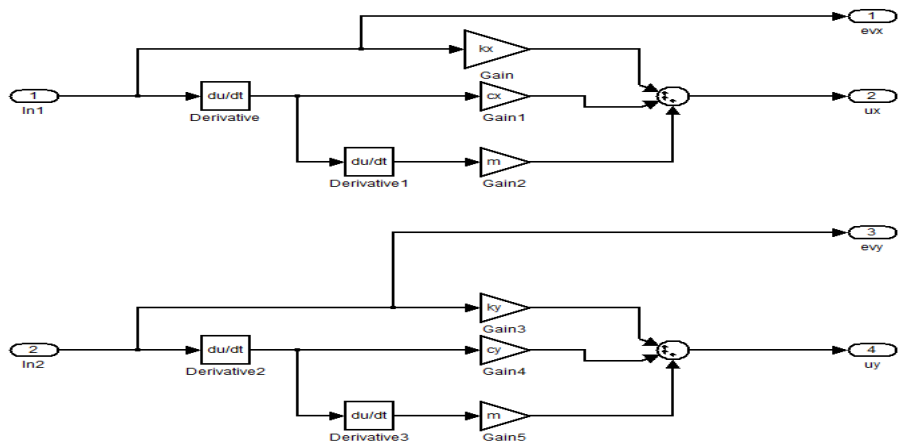

**Figure 4.** Driving force block

The selected delays block (Figure 5 and Figure 6) can help you choose the transmission delays of the relative position and the velocity information between the evader and the chauffeur. The delayed information is used by the controller of the pursuer as if they were fresh.

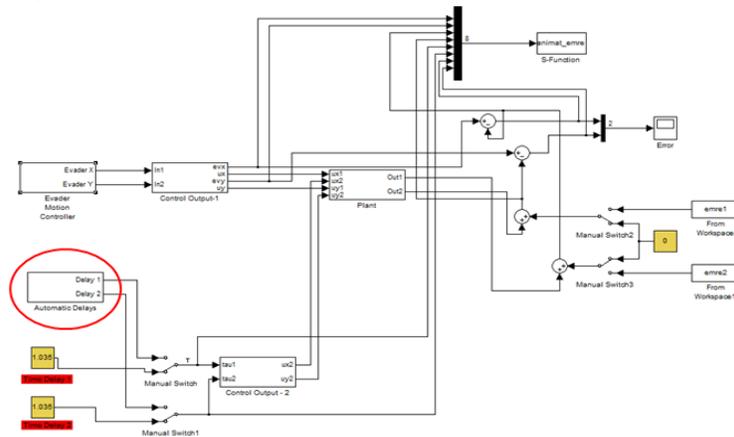

**Figure 5.** Overall plant dynamics with highlighted blocks with automatic delays

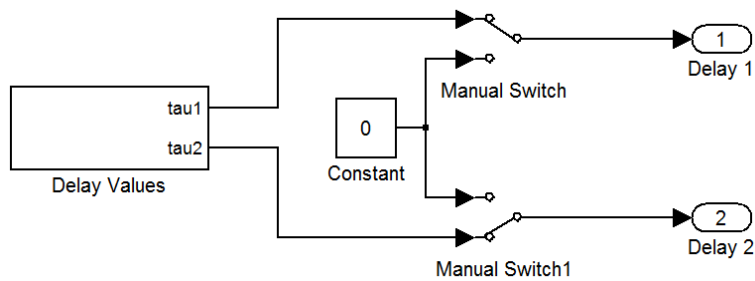

**Figure 6.** Automatic Delays

By double clicking the manual switch, the value of time delay 1 and time delay 2 can be 0 or tau1 and tau2 respectively. Delay values block is shown in the Figure 7.

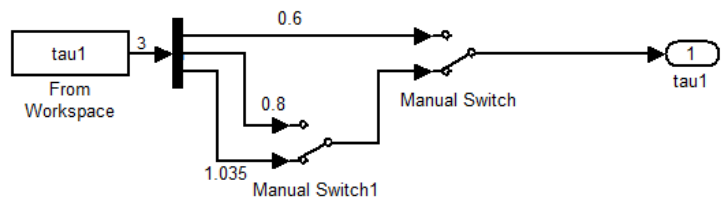

**Figure 7.** Delay Values

Three values 0.6, 0.8 and 1.035 can be chosen automatically. According to the stability map, they are corresponding to the unstable region, stable region, and critical stable region respectively. Time delay values can also be chosen manually as well using the constant blocks (see Figure 5).

Next, the control module for the pursuer is illustrated in Figure 8. The overall system is a PD-type delayed feedback control that deals with the error dynamics as shown in Figure 9.

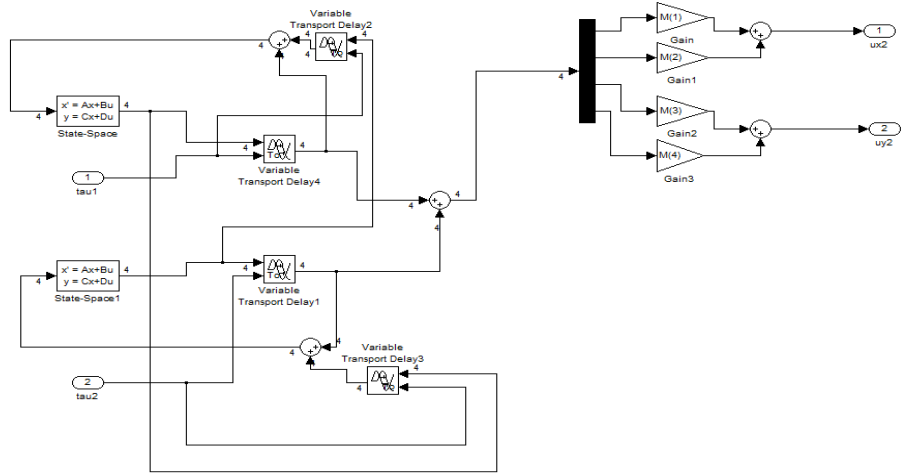

**Figure 8**. Time-Delayed control module for the pursuer

$$\dot{e}(t) = \begin{bmatrix} 0 & 1 & 0 & 0 \\ -305 & -279 & -1 & 1 \\ 0 & 0 & 0 & 1 \\ -1 & 1 & -40 & -2 \end{bmatrix} e(t) + \begin{bmatrix} 0 & 0 & 0 & 0 \\ -14.514 & -1.285 & 22.941 & -0.283 \\ 0 & 0 & 0 & 0 \\ 0 & 0 & 0 & 0 \end{bmatrix} e(t-\tau_1)$$

$$+ \begin{bmatrix} 0 & 0 & 0 & 0 \\ 0 & 0 & 0 & 0 \\ 0 & 0 & 0 & 0 \\ -14 & -1 & 23 & -0.283 \end{bmatrix} e(t-\tau_2)$$

**Figure 9.** Error dynamics of the target tracking game

Since both **C** matrices of the state space models are identity matrices and both **D** matrices are zero, the output of the state space models are the state vectors respectively. The equation for the first state space model is

$$\dot{e}_1 = Ae_1 + B_1 e_1(t-\tau_1) + B_1 e_2(t-\tau_1) \tag{9}$$

and the equation for the second state space model is

$$\dot{e}_2 = Ae_2 + B_2 e_1(t-\tau_2) + B_2 e_2(t-\tau_2) \tag{10}$$

where $e_1$ and $e_2$ are the state vectors of the state space models respectively and $e = e_1 + e_2$, $A, B_1, B_2$ are the matrices in the error dynamics equation (check Figure 9). Summing up these two equations creates the equation (6). The outputs of the time-delayed control are the delayed proportional and derivative control decisions for the pursuer. These two control decisions go into the decoupled 2-D dynamics as shown in Figure 10.

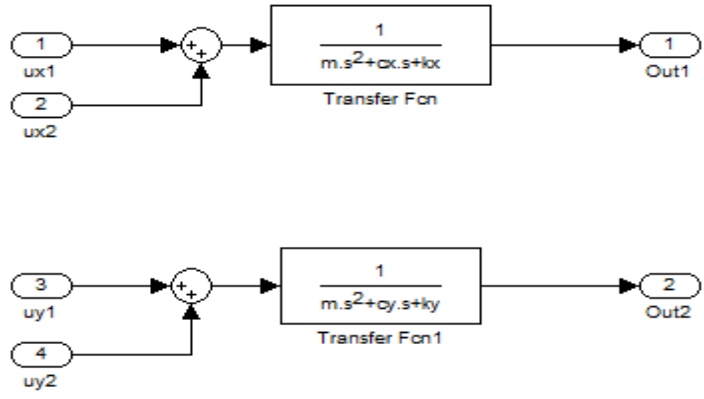

**Figure 10.** Plant

Disturbances in Figure 1 are unexpected and unknown elements which derail the tracking. The feedback control should be able to reject these disturbances. For this, the system should be stable. Figure 11 shows the disturbances to the position of the chauffeur.

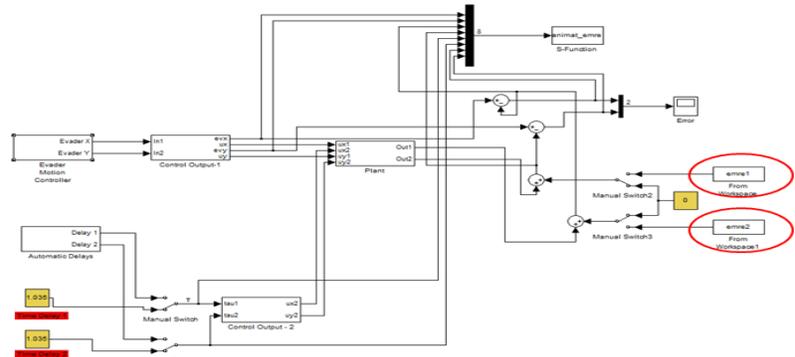

**Figure 11.** Disturbances

The S-Function Block which plots the position of the evader and the pursuer in each sampling time is shown in Figure 12.

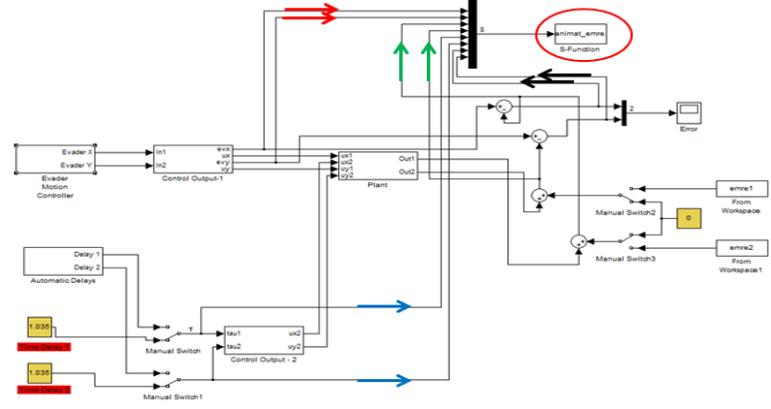

**Figure 12.** S-Function Block

There are eight input signals to the S-Function. First, the two red lines indicate the X and Y position of the evader. Secondly, the two green lines indicate the disturbances.

Next, the two blue lines denote the two delay values in the position measurement and velocity sensing, respectively. Finally, the two black lines are the position error in $x$ and $y$ direction between the pursuer and the evader.

**References**

[1] H. A. P. Blom, E. A. Bloem, and D. Musicki, "JIPDA: Automatic target tracking avoiding track coalescence," *Aerospace and Electronic Systems, IEEE Transactions on,* vol. 51, no. 2, pp. 962-974, 2015.
[2] D. Bresch-Pietri and M. Krstic, "Adaptive trajectory tracking despite unknown input delay and plant parameters," *Automatica,* vol. 45, no. 9, pp. 2074-2081, 2009.
[3] A. Koivo and D. Repperger, "On a game problem involving systems with time delay," *Automatic Control, IEEE Transactions on,* vol. 18, no. 2, pp. 149-152, 1973.
[4] B. M. Mirkin and P.-O. Gutman, "Output-feedback model reference adaptive control for continuous state delay systems.(Author Abstract)," *Journal of Dynamic Systems, Measurement, and Control,* vol. 125, no. 2, p. 257, 2003.
[5] Q. Gao, R. Cepeda-Gomez, and N. Olgac, "The Homicidal Chauffeur Problem with Multiple Time Delayed Feedback," in *Time Delay Systems*, 2012, vol. 10, no. 1, pp. 97-101.
[6] Q. Gao, R. Cepeda-Gomez, and N. Olgac, "A test platform for cognitive delays: target tracking problem with multiple time-delayed feedback control," *International Journal of Dynamics and Control,* vol. 2, no. 1, pp. 77-85, 2014.
[7] R. Ortega and R. Lozano, "Globally stable adaptive controller for systems with delay," *International Journal of Control,* vol. 47, no. 1, pp. 17-23, 1988.
[8] Z. Zhou, M. Zhou, and X. Shi, "Target tracking based on foreground probability," *An International Journal,* vol. 75, no. 6, pp. 3145-3160, 2016.
[9] Z. Zhang and Q. Gao, "Characterization of Extreme Low-Resolution Digital Encoder for Control System with Sinusoidal Reference Signal," in *19th International Conference on Networking, Sensing and Control*, 2017, vol. 11, no. 5, pp. 890-896.
[10] Z. Zhang, N. Olgac, and Q. Gao, "An effective algorithm to achieve accurate sinusoidal amplitude control with a low-resolution encoder," in *Advanced Intelligent Mechatronics (AIM), 2017 IEEE International Conference on*, 2017, pp. 934-939: IEEE.
[11] F. Yao and Q. Gao, "Optimal Control and Switching Mechanism of a Power Station With Identical Generators," in *ASME 2017 Dynamic Systems and Control Conference*, 2017, pp. V001T09A006-V001T09A006: American Society of Mechanical Engineers.
[12] Q. Gao, "Stability Analysis and Novel Control Concepts for Multiple Time-delay Systems," 2015.
[13] Q. Gao and N. Olgac, "Bounds of imaginary spectra of LTI systems in the domain of two of the multiple time delays," *Automatica,* vol. 72, pp. 235-241, 2016.
[14] Q. Gao, Z. Zhang, and C. Yang, "Sign Inverting Control and its Important Properties for Multiple Time-Delayed Systems," in *ASME 2017 Dynamic Systems and Control Conference*, 2017, pp. V001T03A004-V001T03A004: American Society of Mechanical Engineers.
[15] Q. Gao and N. Olgac, "Stability analysis for LTI systems with multiple time delays using the bounds of its imaginary spectra," *Systems & Control Letters,* vol. 102, pp. 112-118, 2017.
[16] Q. Gao and N. Olgac, "Optimal sign inverting control for multiple time-delayed systems," in *American Control Conference (ACC), 2016*, 2016, pp. 979-984: IEEE.
[17] Q. Gao and N. Olgac, "Differentiability of imaginary spectra and determination of its bounds for multiple-delay LTI systems," in *Flexible Automation (ISFA), International Symposium on*, 2016, pp. 296-302: IEEE.
[18] Q. Gao and N. Olgac, "Dixon resultant for cluster treatment of LTI systems with multiple delays," *IFAC-PapersOnLine,* vol. 48, no. 12, pp. 21-26, 2015.
[19] Q. Gao, A. S. Kammer, U. Zalluhoglu, and N. Olgac, "Sign inverting and Delay Scheduling Control concepts with multiple rationally independent delays," in *American Control Conference (ACC), 2014*, 2014, pp. 5546-5551: IEEE.
[20] Q. G. Fulai Yao, "Efficiency optimization of a power station with different generators," in *REM 2017: Renewable Energy integration with Mini/Micro grid Systems, Applied Energy Symposium and Forum 2017*, 2017.

[21] Q. Gao, C. Dong, and H. Jia, "Multiple Time-delay Stability Analysis for the DC- Microgrid Cluster with Distributed Control," in *REM 2017: Renewable Energy integration with Mini/Micro grid Systems, Applied Energy Symposium and Forum 2017*, 2017.

[22] Q. Gao, C. Dong, H. Jia, and Z. Zhang, "Stability Analysis for the Communication Network of DC Microgrids with Cluster Treatment of Characteristic Roots (CTCR) Paradigm," in *REM 2017: Renewable Energy integration with Mini/Micro grid Systems, Applied Energy Symposium and Forum 2017*, 2017.

[23] Y. Huang, Z. Zhang, C. Ji, and D. T. Phan, "Disk drive measuring stroke difference between heads by detecting a difference between ramp contact," ed: Google Patents, 2016.

[24] Z. Zhang, J. Diaz, and N. Olgac, "Adaptive Hybrid Control for Rotationally Oscillating Drill (Ros-Drill©), Using a Low-Resolution Sensor," in *ASME 2011 Dynamic Systems and Control Conference and Bath/ASME Symposium on Fluid Power and Motion Control*, 2011, pp. 463-470: American Society of Mechanical Engineers.

[25] Z. Zhang, *Control Design and Analysis for Rotationally Oscillating Drill (Ros-Drill), with Low-Resolution Feedback*. University of Connecticut, 2012.

[26] Z. Zhang, J. F. Diaz, and N. Olgac, "Adaptive gain scheduling for rotationally oscillating drill, with low-resolution feedback," *International Journal of Mechatronics and Manufacturing Systems,* vol. 6, no. 5-6, pp. 397-421, 2013.

[27] Z. Zhang and N. Olgac, "An Adaptive Control Method for Ros-Drill Cellular Microinjector with Low-Resolution Encoder," *Journal of medical engineering,* vol. 2013, 2013.

[28] Z. Zhang and N. Olgac, "An Adaptive Control Method With Low-Resolution Encoder," in *ASME 2013 Dynamic Systems and Control Conference*, 2013, pp. V003T35A001-V003T35A001: American Society of Mechanical Engineers.

[29] Z. Zhang and N. Olgac, "Zero Magnitude Error Tracking Control for Servo System with Extremely Low-resolution Digital Encoder," *International Journal of Mechatronics and Manufacturing Systems,* vol. 10, no. 4, pp. 355-373, 2017.

[30] Q. Gao, U. Zalluhoglu, and N. Olgac, "Equivalency of Stability Transitions Between the SDS (Spectral Delay Space) and DS (Delay Space)," in *ASME 2013 Dynamic Systems and Control Conference*, 2013, pp. V002T21A001-V002T21A001: American Society of Mechanical Engineers.

[31] Q. Gao, U. Zalluhoglu, and N. Olgac, "Investigation of Local Stability Transitions in the Spectral Delay Space and Delay Space," *Journal of Dynamic Systems, Measurement, and Control,* vol. 136, no. 5, p. 051011, 2014.

[32] K. Gu and M. Naghnaeian, "Stability crossing set for systems with three delays," *Automatic Control, IEEE Transactions on,* vol. 56, no. 1, pp. 11-26, 2011.

[33] J. Chen, G. Gu, and C. N. Nett, "A new method for computing delay margins for stability of linear delay systems," in *Decision and Control, 1994., Proceedings of the 33rd IEEE Conference on*, 1994, vol. 1, pp. 433-437: IEEE.

[34] A. Thowsen, "An analytic stability test for a class of time-delay systems," *Automatic Control, IEEE Transactions on,* vol. 26, no. 3, pp. 735-736, 1981.

[35] Q. Gao, A. S. Kammer, U. Zalluhoglu, and N. Olgac, "Some critical properties of sign inverting control for LTI systems with multiple delays," in *Decision and Control (CDC), 2014 IEEE 53rd Annual Conference on*, 2014, pp. 3976-3981: IEEE.

[36] Q. Gao, A. S. Kammer, U. Zalluhoglu, and N. Olgac, "Combination of sign inverting and delay scheduling control concepts for multiple-delay dynamics," *Systems & Control Letters,* vol. 77, pp. 55-62, 2015.

[37] Q. Gao and N. Olgac, "Optimal sign inverting control for time-delayed systems, a concept study with experiments," *International Journal of Control,* vol. 88, no. 1, pp. 113-122, 2015.

[38] R. Schmid, Q. Gao, and N. Olgac, "Eigenvalue assignment for systems with multiple time-delays," in *2015 Proceedings of the Conference on Control and its Applications*, 2015, pp. 146-152: Society for Industrial and Applied Mathematics.

[39] F. Yao and Q. Gao, "Optimal Control and Switch in a Hydraulic Power Station," in *IEEE IAEAC 2017, Proceedings of 2017 IEEE 2nd Advanced Information Technology, Electronic and Automation Control Conference*, 2017.